\documentclass{DISproc}

\begin{document}
\title{Update and Comparison of Nuclear Parton Distribution Functions and Neutrino DIS.}

\author{{\slshape K.Kova\v{r}\'{\i}k}\\[1ex]
Institute for Theoretical Physics, Karlsruhe Institute of Technology, Karlsruhe, 76128,Germany}

\contribID{250}

\doi  

\maketitle

\begin{abstract}
	We compare the nuclear corrections factors from neutrino deep-inelastic scattering (DIS) with the ones coming from
	a standard analysis of nuclear parton distribution functions (nPDF). We focus on a discrepancy 
	between the most precise neutrino DIS data from NuTeV and the nuclear PDF coming from the analysis 
	of charged lepton DIS and Drell-Yan data.
\end{abstract}

\section{Introduction}
An indispensable part of any prediction for a process measured at a hadron collider such as the LHC are the
parton distribution functions (PDFs). Because of the importance of PDFs, many groups 
perform and update global analyses of PDFs for protons \cite{Ball:2009mk, 
Martin:2009iq, Nadolsky:2008zw} and for nuclei \cite{Hirai:2007sx, 
Eskola:2009uj,deFlorian:2011fp}. 
Proton PDF are determined from data taken not only on protons but from some data taken on nuclear targets, mainly deuterium 
but also heavy nuclei such as lead and iron in case of neutrino DIS. 
Neutrino DIS data is sensitive to the strange quark content of the proton and complements newly available LHC data 
from $W$- or $Z$-boson production.

In order to include the neutrino DIS data in a global fit to help constrain the proton PDF, we 
have to apply a nuclear correction factor. The nuclear correction factor can be obtained either from a 
specific model of nuclear interactions or from an analysis of nuclear parton 
distribution functions (NPDF) based on experimental data.

Here, we discuss a compatibility of neutrino DIS data with the nuclear correction factors obtained from NPDF analysis
focusing on the neutrino DIS data from the NuTeV experiment.
\section{Nuclear correction factors from nuclear PDF}
Nuclear correction factors are in general defined as a ratio of an observable in a nuclear process and the same 
observable in a process involving protons. In the following, we discuss two nuclear correction factors both 
related either to the $F_2$ structure function in neutrino DIS
\begin{equation}
R_{CC}^\nu(F_2;x,Q^2)\simeq \frac{d^A+\bar{u}^A+\ldots}{d^{A,0}+\bar{u}^{A,0}+\ldots}\,,
\label{eq:rcc}
\end{equation}
or to the $F_2$ structure function in charged lepton DIS
\begin{eqnarray}
&& R_{NC}^{e,\mu}(F_2;x,Q^2)\simeq
\frac{[d^A + \bar{d}^A + \ldots]+ 4 [u^A + \bar{u}^A+\ldots]}{[d^{A,0} + \bar{d}^{A,0} + \ldots]
+4 [u^{A,0} + \bar{u}^{A,0}+\ldots]}\,.
\label{eq:rnc}
\end{eqnarray}
The superscript $`0'$ stands for using the free nucleon PDFs $f_i^{p,n}(x,Q)$ as given below in Eq.\ (\ref{eq:pdf}). 

Nuclear correction factors such as those defined by Eqs.~\ref{eq:rcc},\ref{eq:rnc} can be either extracted from the data 
or calculated using the extracted parton distribution functions. Here we use the nuclear PDF from \cite{Schienbein:2009kk}
and \cite{Kovarik:2010uv} where the parameterizations of the nuclear parton distributions of partons in bound protons 
at the input scale of $Q_0=1.3$ GeV are	
\begin{eqnarray}
x\, f_{k}(x,Q_{0}) &=& c_{0}x^{c_{1}}(1-x)^{c_{2}}e^{c_{3}x}(1+e^{c_{4}}x)^{c_{5}}\,,\\ \nonumber
\bar{d}(x,Q_{0})/\bar{u}(x,Q_{0}) &=& c_{0}x^{c_{1}}(1-x)^{c_{2}}+(1+c_{3}x)(1-x)^{c_{4}}\,,
\end{eqnarray}
where $f_k=u_{v},d_{v},g,\bar{u}+\bar{d},s,\bar{s}$ and $\bar{u},\bar{d}$
are a generalization of the parton parameterizations in free protons used in the CTEQ proton analysis \cite{Pumplin:2002vw}. 
To account for different nuclear targets, the coefficients $c_k$ are made to be functions of the nucleon number $A$
\begin{equation}
c_{k}\to c_{k}(A)\equiv c_{k,0}+c_{k,1}\left(1-A^{-c_{k,2}}\right),\ k=\{1,\ldots,5\}\,.\label{eq:Adep}
\end{equation}
From the input distributions, we can construct the PDFs for a general $(A,Z)$-nucleus 
\begin{equation}
f_{i}^{(A,Z)}(x,Q)=\frac{Z}{A}\ f_{i}^{p/A}(x,Q)+\frac{(A-Z)}{A}\ f_{i}^{n/A}(x,Q),\label{eq:pdf}
\end{equation}
where we relate the distributions of a bound neutron, $f_{i}^{n/A}(x,Q)$, to those of a proton by isospin symmetry. 

In the analysis, the same standard kinematic cuts $Q>2\ {\rm GeV}$ and $W>3.5\ {\rm GeV}$ were applied as in 
\cite{Pumplin:2002vw} and we obtain a fit with $\chi^{2}/{\rm dof}$ of 0.946 to 708 data points with 32 free parameters (for 
further details see \cite{Schienbein:2009kk}). 
\begin{figure}[htb]
  \centering
  \includegraphics[width=0.49\textwidth]{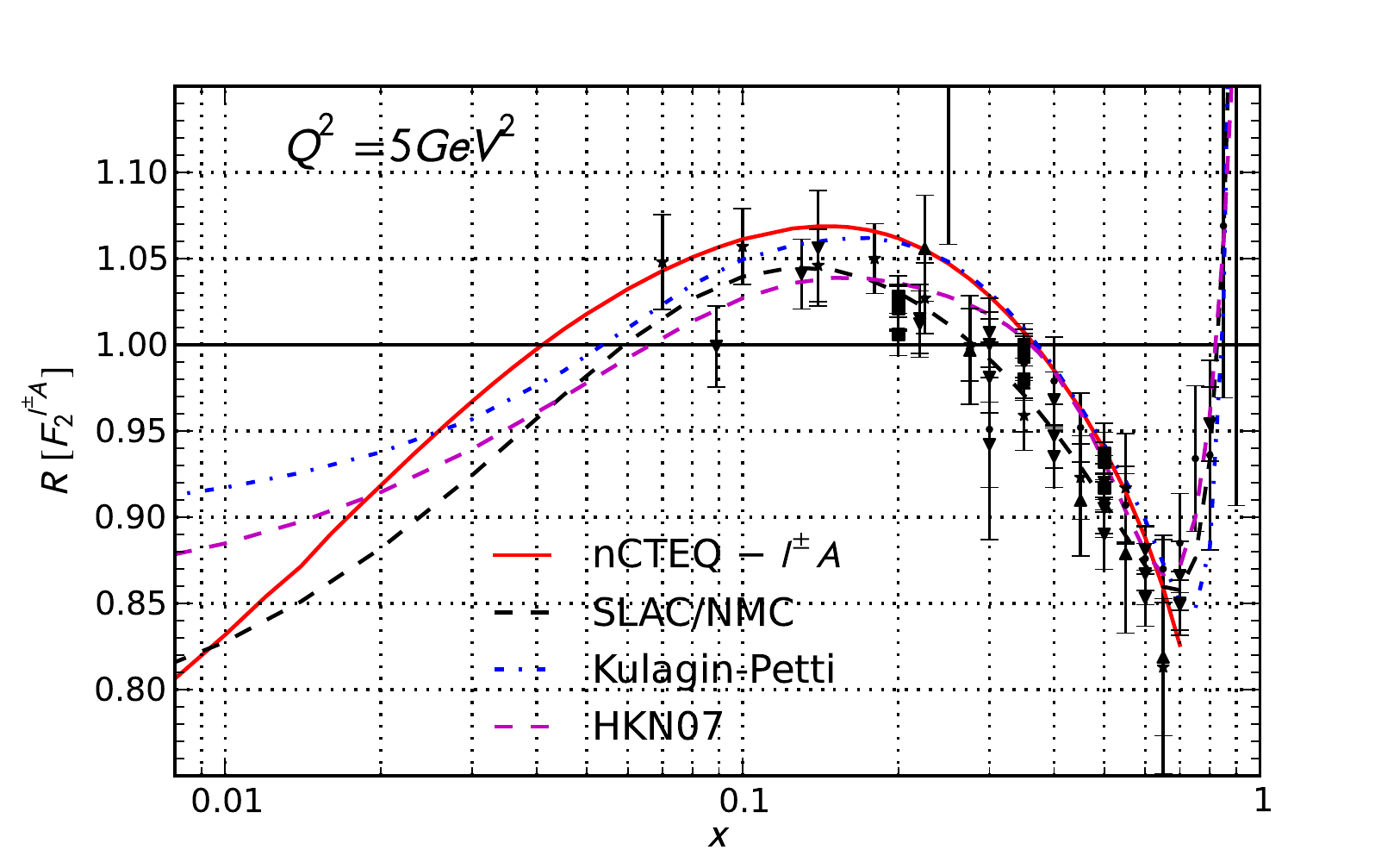}
  \includegraphics[width=0.49\textwidth]{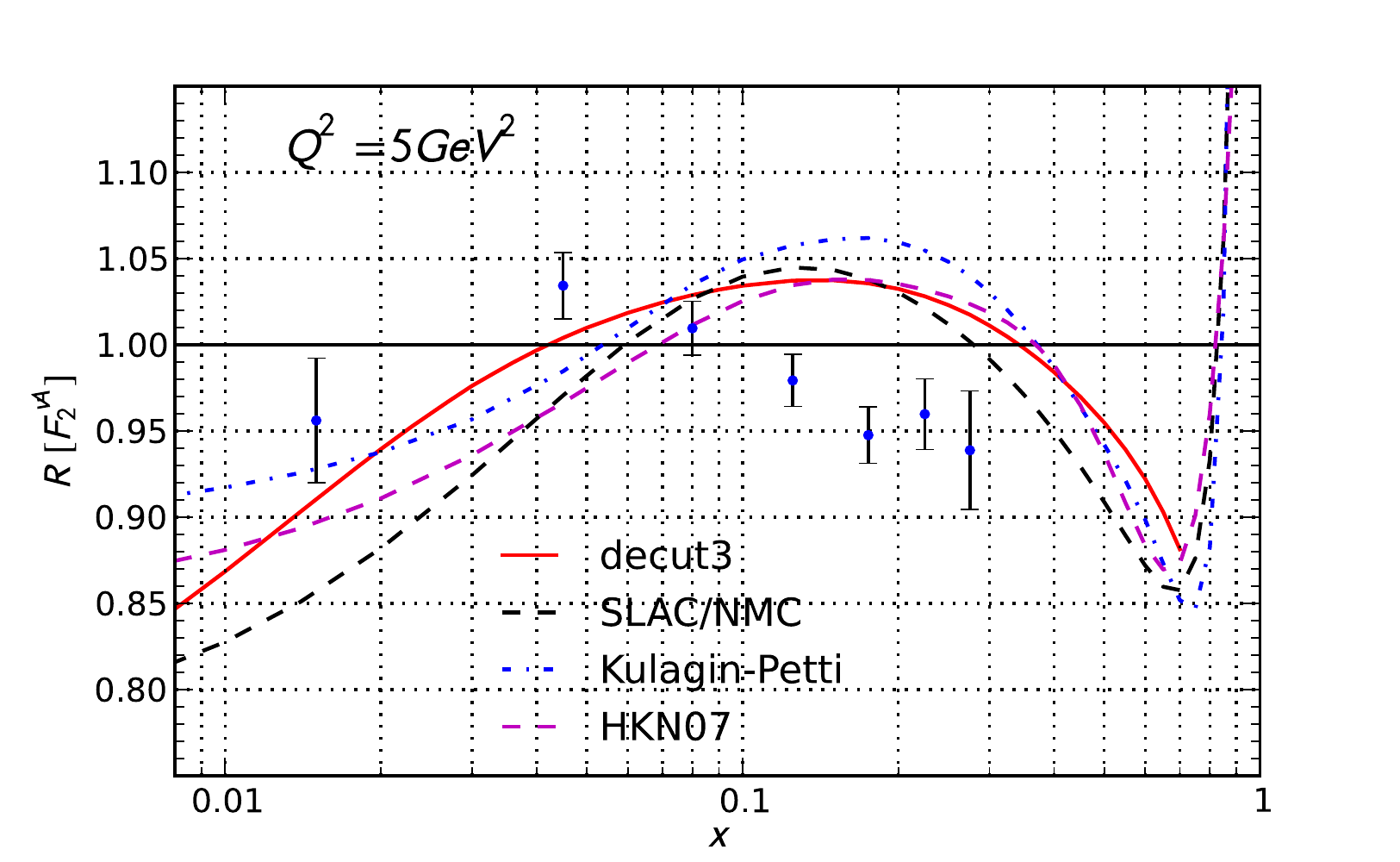}
  \caption{Nuclear correction factors $R_{NC}^{e,\mu}(F_2;x,Q^2)$ and $R_{CC}^\nu(F_2;x,Q^2)$ from the global NPDF analysis compared with the corresponding data for iron target.}\label{Rnc:fig}
\end{figure}
In Fig.~\ref{Rnc:fig} (solid line), we show how the result of our global analysis of charged lepton data translates into 
nuclear correction factors and how the nuclear correction factors compare with experimental data. As first observed in \cite{Schienbein:2007fs}, the 
$R_{CC}^\nu(F_2;x,Q^2)$ correction factor calculated using Eq.~\ref{eq:rcc} with parton densities from the fit to the charged 
lepton nuclear data, does not describe the NuTeV data well which raises the question if including neutrino 
DIS data in the global analysis corrects this behavior without spoiling the $R_{NC}^{e,\mu}(F_2;x,Q^2)$ correction factor which 
fits the charged lepton DIS and DY data well.
\section{Neutrino DIS}
\begin{figure}[htb]
  \centering
  \includegraphics[width=0.49\textwidth]{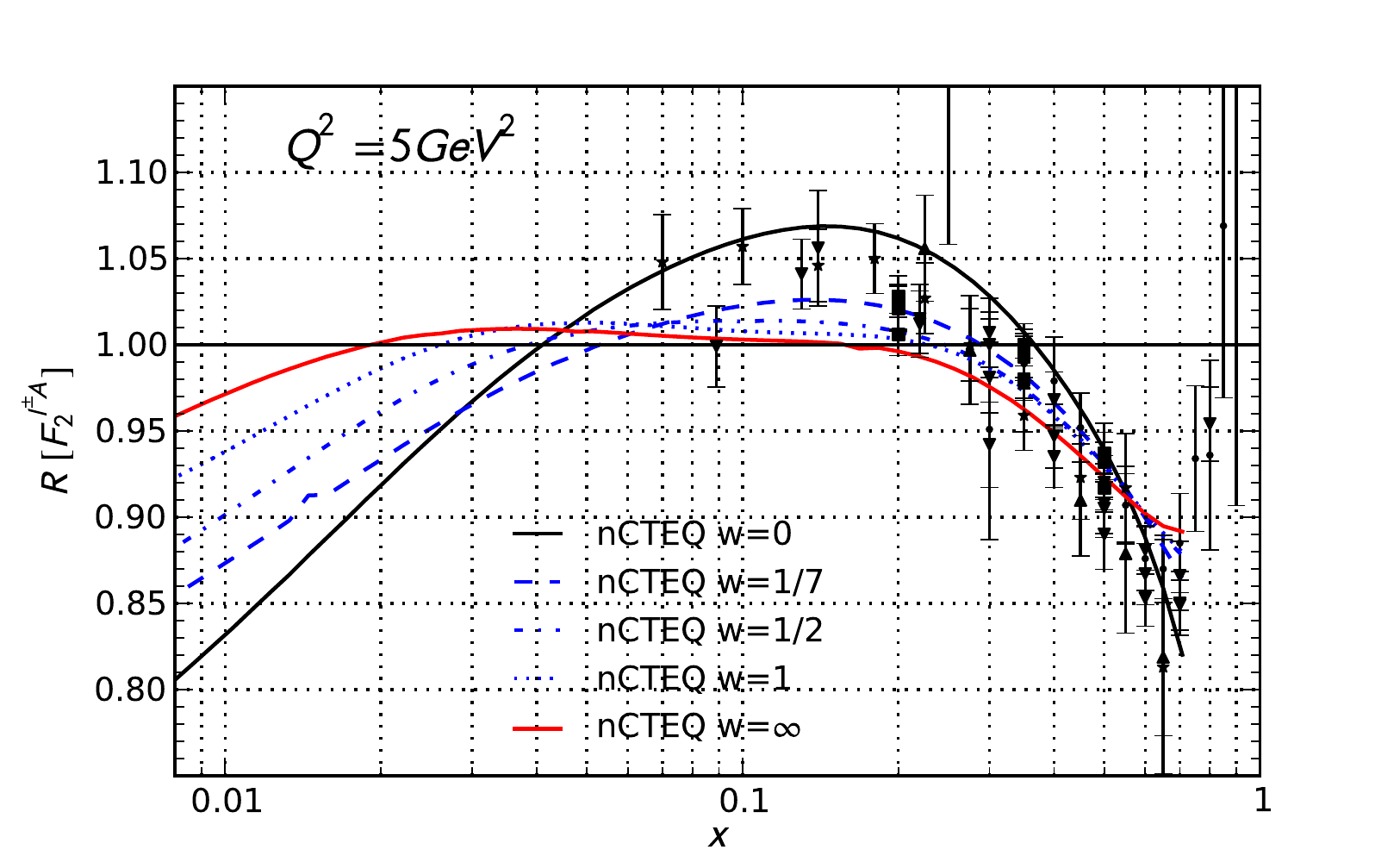}
  \includegraphics[width=0.49\textwidth]{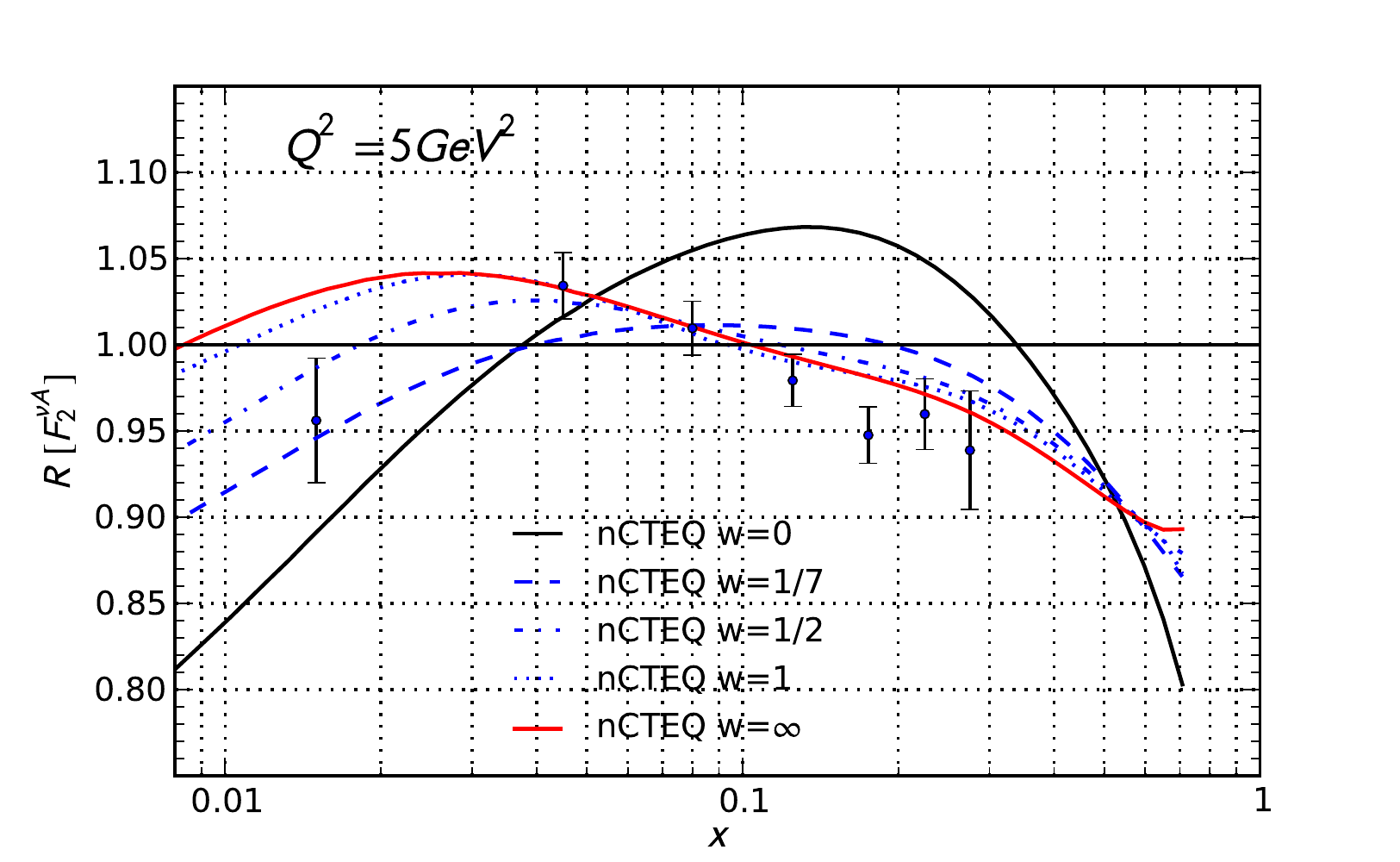}
  \caption{Nuclear correction factors $R_{NC}^{e,\mu}(F_2;x,Q^2)$ and $R_{CC}^\nu(F_2;x,Q^2)$ for all fits in Tab.~\ref{tab:compr}.}\label{weight:fig}
\end{figure}

To analyze the possible discrepancy between the nuclear correction factor $R_{CC}^\nu(F_2;x,Q^2)$ from the 
fit to charged lepton data and the neutrino DIS data, we have included the NuTeV and Chorus neutrino DIS cross-section data 
in the global fit. 
The 3134 neutrino DIS cross-section data points would clearly dominate 708 charged lepton data in the global fit. That is why, 
we have introduced the weight to the neutrino data and set up a series of fits in order to find a compromise fit. 
$\chi^{2}/{\rm dof}$ for each compromise fit with a different weight of neutrino DIS data is listed in Tab.~\ref{tab:compr}. 
Each global fit with a different weight results in a different nuclear correction factor and in Fig.~\ref{weight:fig}
we see that the weight is a suitable parameter which interpolates between the fit using only charged lepton data and 
the fit using only neutrino data (for further details see \cite{Kovarik:2010uv}).
\begin{table}
	\centering
\begin{tabular}{lccc}
\hline 
$w$  & $ \chi^{2}_{l^{\pm}A}$ (/pt)  & $\chi^{2}_{\nu A}$ (/pt)  & total $\chi^{2}$(/pt)\tabularnewline
\hline 
$0$    & 638 (0.90)   & -  & 638 (0.90) \tabularnewline
$1/7$   & 645 (0.91)    & 4710 (1.50)  & 5355 (1.39) \tabularnewline
$1/2$    & 680 (0.96)   & 4405 (1.40)  & 5085 (1.32) \tabularnewline
$1$   & 736 (1.04)      & 4277 (1.36)  & 5014 (1.30) \tabularnewline
$\infty$  & -   & 4192 (1.33)  & 4192 (1.33) \tabularnewline
\hline
\end{tabular}
\caption{Summary table of a family of compromise fits. \label{tab:compr} }
\end{table}
In order to decide on how well the compromise fits describe the data we use the $\chi^2$ goodness-of-fit criterion used in 
\cite{Stump:2001gu,Martin:2009iq}. We consider a fit a good compromise if its $\chi^2$ for both data subsets, the charged lepton DIS and DY 
data and the neutrino DIS data, is within 90\% confidence level of the fits to only charged lepton or neutrino data. 

We define the 90\% percentile $\xi_{90}$ used to define the 90\% confidence level, by
\begin{equation}\label{xi90}
	\int_0^{\xi_{90}}P(\chi^2,N)d\chi^2 = 0.90\,,
\end{equation}
where $N$ is the number of degrees of freedom and $P(\chi^2, N)=\frac{(\chi^2)^{N/2-1}e^{-\chi^2/2}}{2^{N/2}\Gamma(N/2)}$ is the
probability distribution. We can assign a 90\% confidence level error band to the $\chi^2$ of the fits to the charged lepton DIS 
and DY data and to the neutrino DIS data. 
By looking at the overall $\chi^{2}/{\rm dof}$ values and at the plots in
Fig.~\ref{weight:fig}, one might conclude that the global fit with 
$w=1/2$ describes both data well to constitute a compromise fit. The conclusion changes however when we inspect separate 
contributions to the $\chi^{2}/{\rm dof}$ from different experiments. The change in global $\chi^{2}$ is mostly due to change in 
$\chi^{2}$ of the DIS scattering on iron which makes all the compromise fits incompatible.
\begin{figure}[htb]
  \centering
  \includegraphics[width=0.49\textwidth]{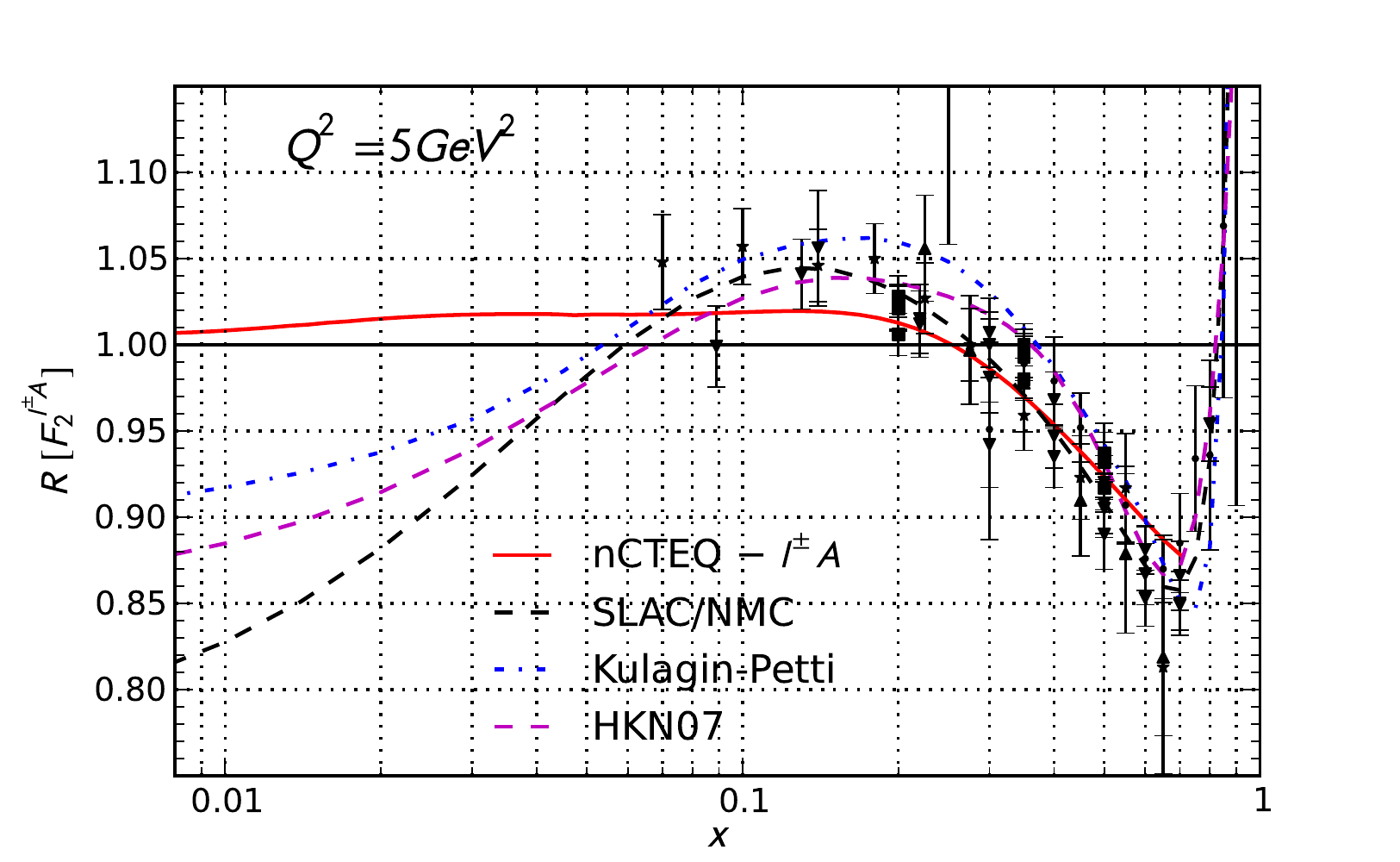}
  \includegraphics[width=0.49\textwidth]{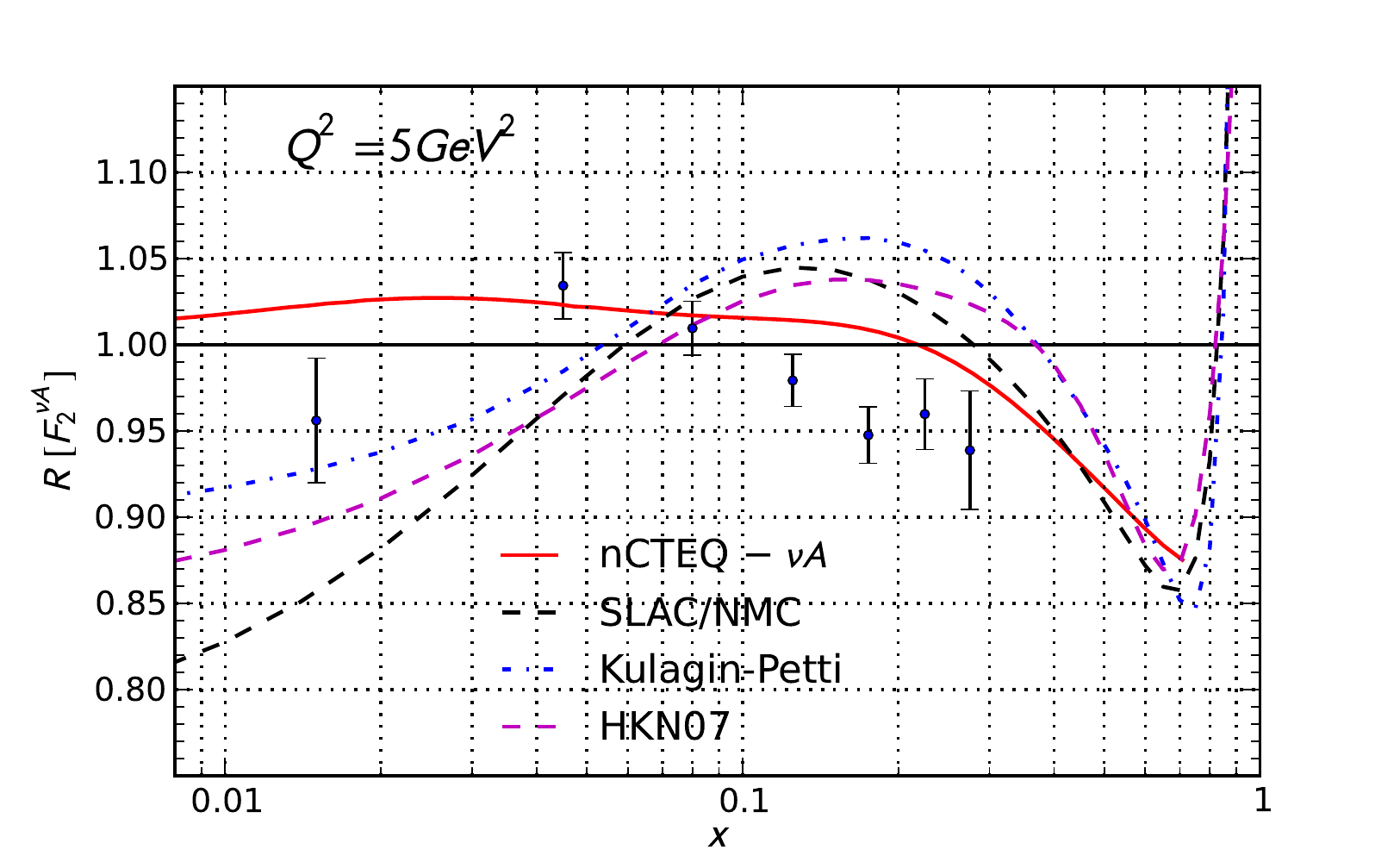}
  \caption{Nuclear correction factors $R_{NC}^{e,\mu}(F_2;x,Q^2)$ and $R_{CC}^\nu(F_2;x,Q^2)$ where neutrino data were
included with uncorrelated systematic errors.}\label{uncorr:fig}
\end{figure}
The conclusion about incompatibility of the fits rests on NuTeV data having small errors which can be demonstrated by 
neglecting the correlations in systematic errors which results in a compatible fit of all the data (see Fig.~\ref{uncorr:fig}).
\section{Conclusion}
A thorough global NPDF analysis of the combined charged lepton and neutrino data leads us to conclude that there is 
no good compromise description of both the data sets simultaneously. The differences can be seen in the low and 
intermediate $x$ regions where the neutrino DIS (NuTeV) do not show a strong shadowing effect as the charged lepton data 
do. The inability to describe all data by one consistent framework poses problems for including the NuTeV data to the proton
PDF analysis.
%

{\raggedright
\begin{footnotesize}


 \bibliographystyle{DISproc}
 \bibliography{kovarik_karol_npdf.bib}

\providecommand{\href}[2]{#2}\begingroup\raggedright\begin{thebibliography}{10}

\bibitem{Ball:2009mk}
R.~D. Ball {\em et~al.}
\newblock \href{http://dx.doi.org/10.1016/j.nuclphysb.2009.08.003}{Nucl. Phys.
  {\bfseries B823} (2009) 195--233},
\href{http://arxiv.org/abs/0906.1958}{{\ttfamily arXiv:0906.1958 [hep-ph]}}.

\bibitem{Martin:2009iq}
A.~D. Martin, W.~J. Stirling, R.~S. Thorne, and G.~Watt.
\newblock \href{http://dx.doi.org/10.1140/epjc/s10052-009-1072-5}{Eur. Phys. J.
  {\bfseries C63} (2009) 189--285},
\href{http://arxiv.org/abs/0901.0002}{{\ttfamily arXiv:0901.0002 [hep-ph]}}.

\bibitem{Nadolsky:2008zw}
P.~M. Nadolsky {\em et~al.}
\newblock \href{http://dx.doi.org/10.1103/PhysRevD.78.013004}{Phys. Rev.
  {\bfseries D78} (2008) 013004},
\href{http://arxiv.org/abs/0802.0007}{{\ttfamily arXiv:0802.0007 [hep-ph]}}.

\bibitem{Hirai:2007sx}
M.~Hirai, S.~Kumano, and T.~H. Nagai.
\newblock \href{http://dx.doi.org/10.1103/PhysRevC.76.065207}{Phys. Rev.
  {\bfseries C76} (2007) 065207},
\href{http://arxiv.org/abs/0709.3038}{{\ttfamily arXiv:0709.3038 [hep-ph]}}.

\bibitem{Eskola:2009uj}
K.~J. Eskola, H.~Paukkunen, and C.~A. Salgado.
\newblock \href{http://dx.doi.org/10.1088/1126-6708/2009/04/065}{JHEP
  {\bfseries 04} (2009) 065},
\href{http://arxiv.org/abs/0902.4154}{{\ttfamily arXiv:0902.4154 [hep-ph]}}.

\bibitem{deFlorian:2011fp}
D.~de~Florian, R.~Sassot, P.~Zurita, and M.~Stratmann.
\newblock Phys.Rev. {\bfseries D85} (2012) 074028,
\href{http://arxiv.org/abs/1112.6324}{{\ttfamily arXiv:1112.6324 [hep-ph]}}.

\bibitem{Schienbein:2009kk}
I.~Schienbein {\em et~al.}
\newblock \href{http://dx.doi.org/10.1103/PhysRevD.80.094004}{Phys. Rev.
  {\bfseries D80} (2009) 094004},
\href{http://arxiv.org/abs/0907.2357}{{\ttfamily arXiv:0907.2357 [hep-ph]}}.

\bibitem{Kovarik:2010uv}
K.~Kova\v{r}\'{\i}k {\em et~al.}
\newblock
  \href{http://dx.doi.org/10.1103/PhysRevLett.106.122301}{Phys.Rev.Lett.
  {\bfseries 106} (2010) 122301},
\href{http://arxiv.org/abs/1012.0286}{{\ttfamily arXiv:1012.0286 [hep-ph]}}.

\bibitem{Pumplin:2002vw}
J.~Pumplin {\em et~al.}
\newblock JHEP {\bfseries 07} (2002) 012,
\href{http://arxiv.org/abs/hep-ph/0201195}{{\ttfamily hep-ph/0201195}}.

\bibitem{Schienbein:2007fs}
I.~Schienbein {\em et~al.}
\newblock \href{http://dx.doi.org/10.1103/PhysRevD.77.054013}{Phys. Rev.
  {\bfseries D77} (2008) 054013},
\href{http://arxiv.org/abs/0710.4897}{{\ttfamily arXiv:0710.4897 [hep-ph]}}.

\bibitem{Stump:2001gu}
D.~Stump {\em et~al.}
\newblock \href{http://dx.doi.org/10.1103/PhysRevD.65.014012}{Phys. Rev.
  {\bfseries D65} (2001) 014012},
\href{http://arxiv.org/abs/hep-ph/0101051}{{\ttfamily arXiv:hep-ph/0101051}}.

\end{thebibliography}\endgroup
\end{footnotesize}
}


\end{document}